# NMR evidence for selective enhancement of Mo magnetic moment by electron doping in $Sr_xLa_{2-x}FeMoO_6$

M. Wojcik, E. Jedryka, S. Nadolski

*Institute of Physics, Polish Academy of Sciences, Al. Lotników 32/46, 02 668 Warszawa, Poland*

J. Navarro, D.Rubi and J. Fontcuberta

*Institut de Ciencia de Materials de Barcelona, CSIC, Campus Universitari de Bellatera, E-08193 Bellatera, Spain.*

## ABSTRACT

$^{95,\ 97}$Mo NMR experiments have been performed on a series of $Sr_2FeMoO_6$ and electron-doped $Sr_{2-x}La_xFeMoO_6$ ceramics. Detailed analysis of the NMR spectra from pristine $Sr_2FeMoO_6$ conclusively shows that the Mo hyperfine field is mainly due to atomic Mo magnetic moments. No contribution of transferred hyperfine field has been observed, confirming the absence of s-electrons in the conduction band. Upon La doping, the NMR frequency (hyperfine field) gradually increases proving that the concentration of spin polarized electrons at Mo ion is enhanced by the La substitution. A simple linear correlation between magnetic moment at Mo sites and the Curie temperature of the system has been found. Implications for understanding the electronic structure and the ferromagnetic coupling in these systems are underlined.

Magnetic oxides with the double perovskite structure, derived from the parent compound $Sr_2FeMoO_6$ (SFMO), have recently received a lot of attention following a report of Kobayashi et al. that this material is a half-metallic ferromagnet with the Curie temperature of 415 K [1], making it a very suitable candidate for applications in magnetoelectronics. Its crystal structure (denoted as $A_2BB'O_6$) consists of an ordered arrangement of perovskite ($ABO_3$) building units, in such way that their B sites are alternatively occupied by Fe and Mo ions. Sr ions occupy the central A positions in each perovskite unit.

The electronic structure consists of localized *spin-up* states originating from Fe 3d orbitals ($t_{2g}$ and $e_g$) and a conduction band formed by hybridized Fe 3d and Mo 4d *spin-down* $t_{2g}$ states [2] . The *spin-up* states are fully occupied by five 3d electrons of the high-spin state of $Fe^{3+}$, whereas the conducting *spin down* band is partially filled by one d electron formally originating from $Mo^{5+}$. In this way, the experimentally observed ferromagnetic alignment of localized Fe spins (S=5/2) implies that the magnetic polarization of Mo ion is antiparallel with respect to the Fe moments.

However, it is not obvious how magnetic interaction is transmitted. It is experimentally found that magnetic moment at Mo sites is very small (~0.3 $\mu_B$ [3]) thus suggesting a weak exchange with neighbouring Fe-ions. In spite of this, the Curie temperature is well above that of manganites. Consequently, enormous efforts have been dedicated to obtain experimental confirmation of the antiferromagnetic coupling between itinerant carriers and localized moments at Fe-sites and to reach a microscopic understanding of the nature of the magnetic interactions. An important breakthrough was the demonstration (on grounds of the analysis of paramagnetic susceptibility) that the localized Fe moments are antiferromagnetically coupled to the conduction electrons [4]. This antiferromagnetic interaction governs the ferromagnetic order of the Fe sublattice, and consequently the Curie temperature should depend on the strength of the coupling between the conduction electrons in the hybridized Fe and Mo $t_{2g}$ band and the Fe localized moments. It was thus postulated that the Curie temperature can be increased by enhancing electron density in the conduction band, e.g. by partial substitution of divalent Sr with trivalent La. The correctness of this approach has been successfully demonstrated and the systematic increase of $T_c$ upon doping with La was achieved in $Sr_{2-x}La_xFeMoO_6$ [5-9]. Recently, it has also been shown that other electron donors –such as $Nd^{3+}$- when substituted into SFMO also promote an enhancement of the Curie temperature [10]. Subsequent photoemission (PES) studies performed on $Sr_{2-x}La_xFeMoO_6$ have confirmed that indeed density of states at the Fermi edge increases upon La doping [11]. Moreover, the observed increase of density of states seems to arise selectively from the Mo $t_{2g}$ *spin-down* states. However, the PES data cannot positively attribute the observed changes in the density of states to a purely electronic effect, since the La/Sr substitution introduces a modification of the Mo-O-Fe bond topology [9].

Therefore, with the aim to obtain a microscopic information on the carrier injection mechanism, we have performed an extensive Nuclear Magnetic Resonance (NMR) study on a series of $Sr_{2-x}La_xFeMoO_6$ (LSMO) ($0 \leq x \leq 0.8$) samples. In order to separate the effects of electron doping (e-doping) from possible lattice distortion due to modification of the A sublattice, a reference sample isoelectronic doped with $Ca^{2+}$ ions of composition $Sr_{1.8}Ca_{0.2}FeMoO_6$ has also been studied.

We will first show that the $^{95,97}Mo$ NMR spectrum is insensitive to the disorder on either BB' sublattice (Fe-Mo antisites) or AA'sublattice (Sr/La or Sr/Ca distribution). We will argue that this remarkable property is a direct consequence of the electronic structure of the SFMO and the absence of s electrons in the conducting band. This implies that the transferred hyperfine field on Mo is negligible and the measured hyperfine field (HF) is determined only by the *on-site* magnetic moment ($\mu$(Mo)). It follows that, in SFMO, HF is a direct measure of $\mu$(Mo) and thus it can be used as a probe of carrier injection into Mo orbitals. We will show that the NMR spectra of La-substituted LSMO reveal a dramatic increase of $\mu$(Mo) which is thus a clear evidence of the e-doping into Mo orbitals. Moreover, we are able to prove that the hyperfine field on Fe remains invariant with the e-doping. The implication of these findings for the understanding of the ferromagnetic coupling in double perovskites is emphasized.

Three different sets of samples have been used. Undoped SFMO oxides having distinct concentration of antisites (AS) (determined from refinement of X-ray diffraction patterns and magnetization measurements as described in [5, 12]) have been prepared as described in [5,8,12]. The corresponding AS concentrations are: 3%, 7%, 14% and 28% and the corresponding saturation magnetization ($M_S$) values are: 3.95 $\mu_B$, 3.67 $\mu_B$, 2.89 $\mu_B$ and 1.76 $\mu_B$ , respectively. Another set of samples with composition $Sr_{2-x}La_xFeMoO_6$ ($0 \leq x \leq 0.8$), having $T_C$= 432K, 440K, 455K, and 473K for x=0.2, 0.4 and 0.6 and 0.8 respectively, has also been prepared. Refinement of neutron diffraction profiles have been used to confirm that, within the experimental resolution, all samples here reported are oxygen stoichiometric [9]. Finally, a sample of composition $Sr_{1.8}Ca_{0.2}FeMoO_6$ (AS= 6% and $M_S$= 3.6 $\mu_B$) has also been prepared and characterized in a similar manner.

NMR spectra have been recorded in the frequency range from 20 MHz up to 200 MHz, using an automated, coherent, phase sensitive spin-echo spectrometer.

Experiments have been carried out at 4.2 K in zero external magnetic field and varying the power of r.f. pulses. A usual $\omega^2$ correction of signal intensity has been applied as well as a correction for the intrinsic enhancement factor, which was determined at each frequency from the spin echo intensity dependence on the excitation r.f. power level [13].

In Fig.1 we show the $^{95,97}$Mo NMR spectra of SFMO specimens having distinct AS concentrations (3%, 7%, 14% and 28%). The spectrum consists of the main asymmetric resonance line with a peak at about ~67 MHz and a low frequency tail with broad maxima. The macroscopic magnetic parameters of studied samples, such as the NMR restoring field and the magnetic stiffness defined as the saturation magnetization over the initial susceptibility ($1/\chi_{in} * M_{sat}$), are plotted in the inset in Fig.1. We note that their values increase by an order of magnitude over the studied AS concentration range, thus indicating that AS harden the magnetic system [14]. At the same time, the NMR spectra recorded in the frequency range from 20 MHz to 90 MHz are totally insensitive to the degree of B/B' sublattice disorder. This surprising observation can be understood considering the origins of hyperfine field, which can be expressed by the following formula:

$$HF = A_{core}\mu_l + A_{cond}\mu_l + A_{tran}\Sigma n_i \mu_j \qquad (1)$$

The hyperfine coupling coefficients have the following origin: $A_{core}$ - core polarization due to the exchange interaction between s electrons of the inner shells and the on-site magnetic moment of the d electrons, $A_{cond}$ - spin polarization of the s conduction electrons due to the on-site magnetic moment of the atom itself, $A_{tran}$ - conduction s electron polarization due to the moments of neighbouring atoms (transferred hyperfine field).

According to electronic structure calculations of SFMO [1,2] the conduction band is primarily formed by admixtures of 3d(Fe) and 4d(Mo) and 2p(O) orbitals. Due to the absence of significant s electron contributions, the second and the third terms in eqn.1 are inactive and the *on-site* magnetic moment becomes the only source of hyperfine field on Mo, i.e HF ≈ $A_{core}\mu_{l.}$

In the SFMO structure (space groups I4/m or $P2_1/n$ [9]) all Mo sites are equivalent, giving rise to the main resonance line at ~67 MHz. The low frequency tail (Fig.1a), being independent of AS concentration (see Fig. 1a) and present in both ceramic and single crystal specimens [15], can not be attributed to antisites or surface effects. Possible scenarios for the origin of the low frequency tail will be presented later.

In Fig. 2 we show the NMR spectra recorded at 4.2 K from some representative SFMO specimens containing an admixture in the A sublattice. In the middle panel the spectrum recorded from the pristine SFMO (AS= 5 %) sample is shown as a reference. As can be seen in Fig.2a (top) substitution of as much as 10% of Ca in the A sublattice and the concomitant Ca/Sr disorder, does not influence significantly the resonance line position in the NMR spectrum. The only effect of Ca doping is reflected in a weak asymmetric broadening of the main line, which reveals its doublet-like structure: two peaks can now be distinguished, one at 65.8 MHz and the other one at 68 MHz. The peak at 65.8 MHz has a distinctly higher restoring field (NMR enhancement factor), suggesting its different origin with respect to the rest of the spectrum. Indeed, comparing the $^{57}$Fe hyperfine field value given by the Mössbauer experiment: 47.7 T [16] this line can be positively interpreted as the $^{57}$Fe

NMR signal. To illustrate its contribution, we present it as a shaded part of the overall intensity and, additionally, we plot it as a separate shaded line at the bottom of Fig.2a. The remaining intensity in the NMR spectrum is attributed to the $^{95}$Mo and $^{97}$Mo isotopes - the gyromagnetic constants of these Mo isotopes are too close to each other (1.7433, -1.7799 ($10^7$ rad $T^{-1}s^{-1}$) for $^{95}$Mo and $^{97}$Mo , respectively [17]), to separate their respective NMR signals.

In sharp contrast to the mild influence of Ca doping, the admixture of La has a dramatic influence on the NMR spectra: the resonant lines rapidly broaden up and shift towards higher frequencies. Fig. 2c illustrates this effect for x=0.2. Remarkably, the $^{57}$Fe NMR line at 65.8 MHz does not change its position and can now be resolved from the background of a much stronger main Mo NMR signal which is shifted towards higher frequencies.

In the crystal structure of SFMO, every site in the A sublattice has four Mo and four Fe neighbours of the B/B' sublattice. The fact that Fe local HF field is not modified by La substitution confirms that the local environment effects (modification of a local field via transferred hyperfine field) are completely inactive in these materials, as it was already indicated by the insensitivity of hyperfine fields to presence of Fe-Mo antisites. While the magnetic moment on Fe did not change with 10% La substitution, the Mo magnetic moment did increase considerably, as evidenced by the NMR frequency up shift (Fig.1b-bottom). This means that the additional electrons introduced into the system by replacing divalent Sr with trivalent La enhance selectively the charge density at the Mo sites, not altering the charge at the Fe site (at least for low La concentration). This result can be considered as an experimental evidence of the predicted tendency for the additional electrons to enter the Mo orbitals.

Fig. 3 shows the evolution of the NMR spectrum of $Sr_{2-x}La_xFeMoO_6$ for various La content. With the growing La content the resonance line is clearly shifted towards higher frequency. The linewidth increases rapidly and the resolution in the low frequency tail is quickly lost. In order to bring out the details of the increasingly featureless spectra and follow their evolution upon La doping, we show in the inset to Fig. 3 some of the spectra normalized to the same amplitude and fitted with 3 gaussian lines. The frequency of the main NMR line ($\nu_0$) as well as the spectrum gravity center ($<\nu>$) are presented in Fig.4 as a function of La content. From data in Fig. 4 it is clear that the NMR resonance frequency increases roughly linearly with the La concentration. The frequency of the main line increases at a rate $d\nu_0/dx$ of about ~49.5 MHz/(La atom). This means that replacement of one $Sr^{2+}$ ion by one $La^{3+}$ ion in the SFMO lattice, causes an increase by 72% of the hyperfine field at Mo nucleus. The spectrum gravity center moves ($d<\nu>/dx$) upwards with a lower slope of only ~28 MHz/(La atom). The slower frequency up-shift of the spectrum gravity center possibly reflects that the influence of La is not homogenous for all Mo ions. As we have argued above, the Fe/Mo antisites are not expected to be the source of this inhomogeneity.

More important, the frequency up-shift of the Mo NMR spectrum indicates the increasing presence of the additional electrons in the immediate vicinity of Mo site, which are responsible for the increase of magnetic moment and local field on Mo. Assuming that only the $t_{2g}$ band is being filled, the increase of magnetic moment is equivalent to the growing number of “spin down” electrons. We note that this observation is fully consistent with the reported evolution of the unit cell upon e-doping and the tiny expansion of the Mo-O octahedra as observed by neutron diffraction [7,9].

According to Tovar et al [4], the filling up of the conduction band should promote an increase of the Curie temperature of SFMO. Indeed, our study indicates a simple linear relationship with a slope of ~1.8 K/MHz between the average NMR frequency (<ν>) for the particular samples and the corresponding critical temperatures, as shown in Fig. 3b. This is a solid experimental evidence of a direct correlation between the number of electrons on Mo and the Curie temperature of the system. That is: not only the density of states at the Fermi edge projected on Mo orbitals increases as indicated by PES experiments [11] but also the concentration of carriers is clearly enhanced upon La substitution.

Before concluding we would like to indicate that the low-frequency tail in the NMR spectra of the pristine SFMO (Figs. 1a and 1b) could find its origin in an intrinsic electronic phase separation, producing non-equivalent Mo ions. We have also mentioned that low-frequency tail shifts to higher frequency upon La doping but with a smaller slope than the main resonance, accompanied by the visible broadening of the spectra. This is an indication that not only Mo magnetic moment itself, but also the NMR spectrum features (linewidth and mechanisms responsible for the low frequency structure) are proportional to the electronic charge. The charge-sensitive low frequency structure observed in the NMR spectra might be thus regarded as a manifestation of Mo moment instability. Indeed, the existence of some charge separation in double perovskites has been recently proposed by Sarma et al. [2].

In summary, we have shown that La doping promotes a carrier injection (electrons) into the conduction band of $Sr_2FeMoO_6$ that produces an increase of the resonant frequency reflecting the enhancement of the magnetic moment of Mo. These results provide a convincing evidence of selective carrier injection in double perovskites and of its relevance for the strength of the magnetic coupling.

This work has been supported in part by Research Framework Programme V (Growth) of the European Community under contract number G5RD-CT2000-00138" (AMORE), by the KBN grant number 72/E-67/SPUB/5.PR UE/DZ 481/2002-2003, by the grant from the Ford Motor Company (Poland) and by the MCyT (Spain) project MAT2002-03431.

## Figure Captions

**Fig. 1** NMR spectra at 4.2 K of: $Sr_2FeMoO_6$ ceramics having different antisites (AS) concentrations. Inset: NMR restoring field (squares) and sample's magnetic stiffness (triangles) as a function of AS concentration.

**Fig.2** NMR spectra recorded at 4.2 K from (a) isoelectronic substituted ($Sr_{1.8}Ca_{0.2}FeMoO_6$) , (b) $Sr_2FeMoO_6$ (AS= 5 % ) and (c) heterovalent substituted ($Sr_{1.8}La_{0.2}FeMoO_6$) .

**Fig. 3** (a) NMR spectra of $Sr_{2-x}La_xFeMoO_6$ ceramics. (b) Detailed view of some of the spectra showing a deconvolution as described in the text.

**Fig. 4** (a) NMR frequencies of the main line ($\nu_0$) and of the spectrum gravity center ($\langle\nu\rangle$) as a function of the La content (x) in $Sr_{2-x}La_xFeMoO_6$. (b) Relationship between the Curie temperature $T_C$ and $\langle\nu\rangle$.

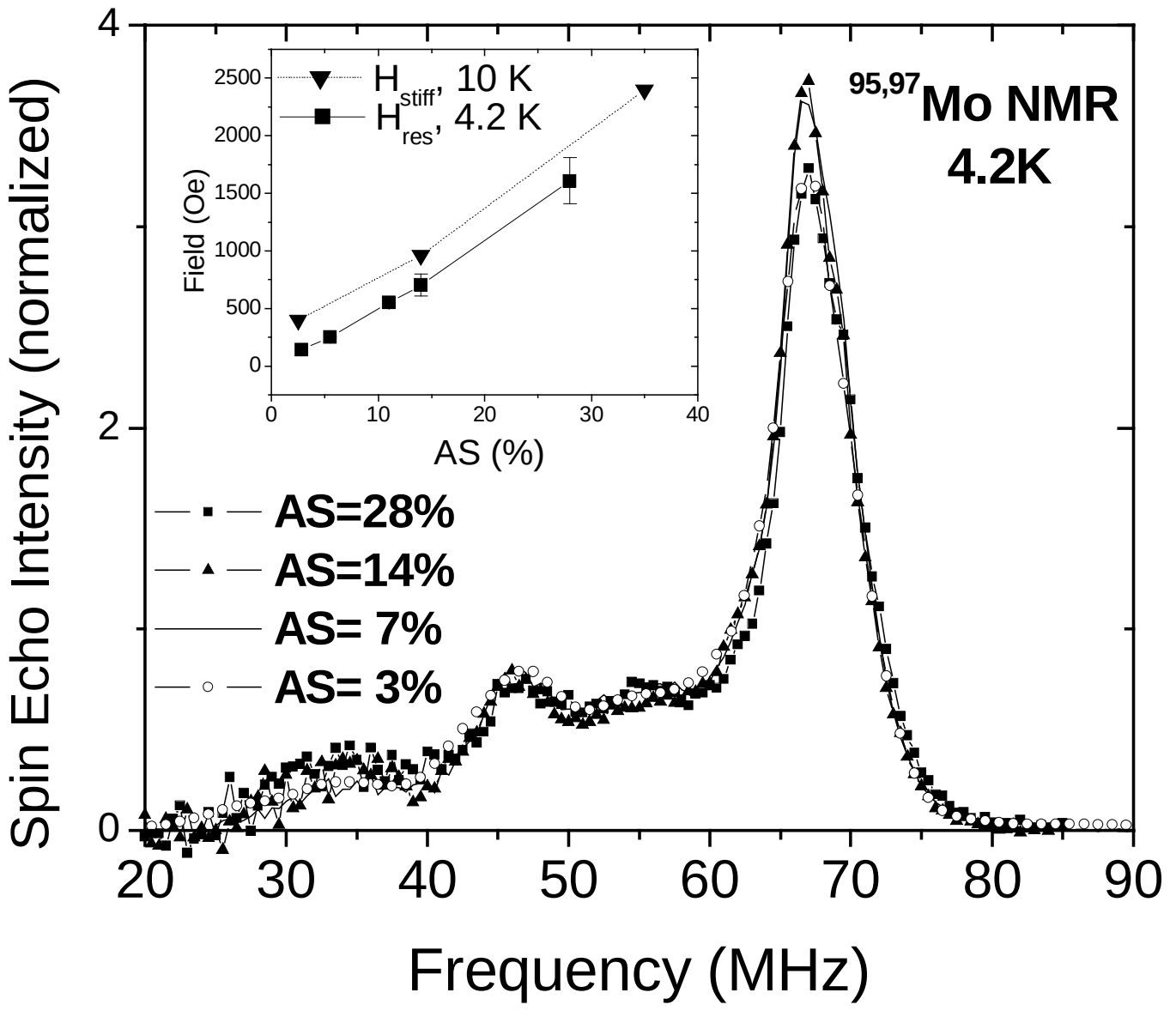

Fig.1, M. Wojcik et al

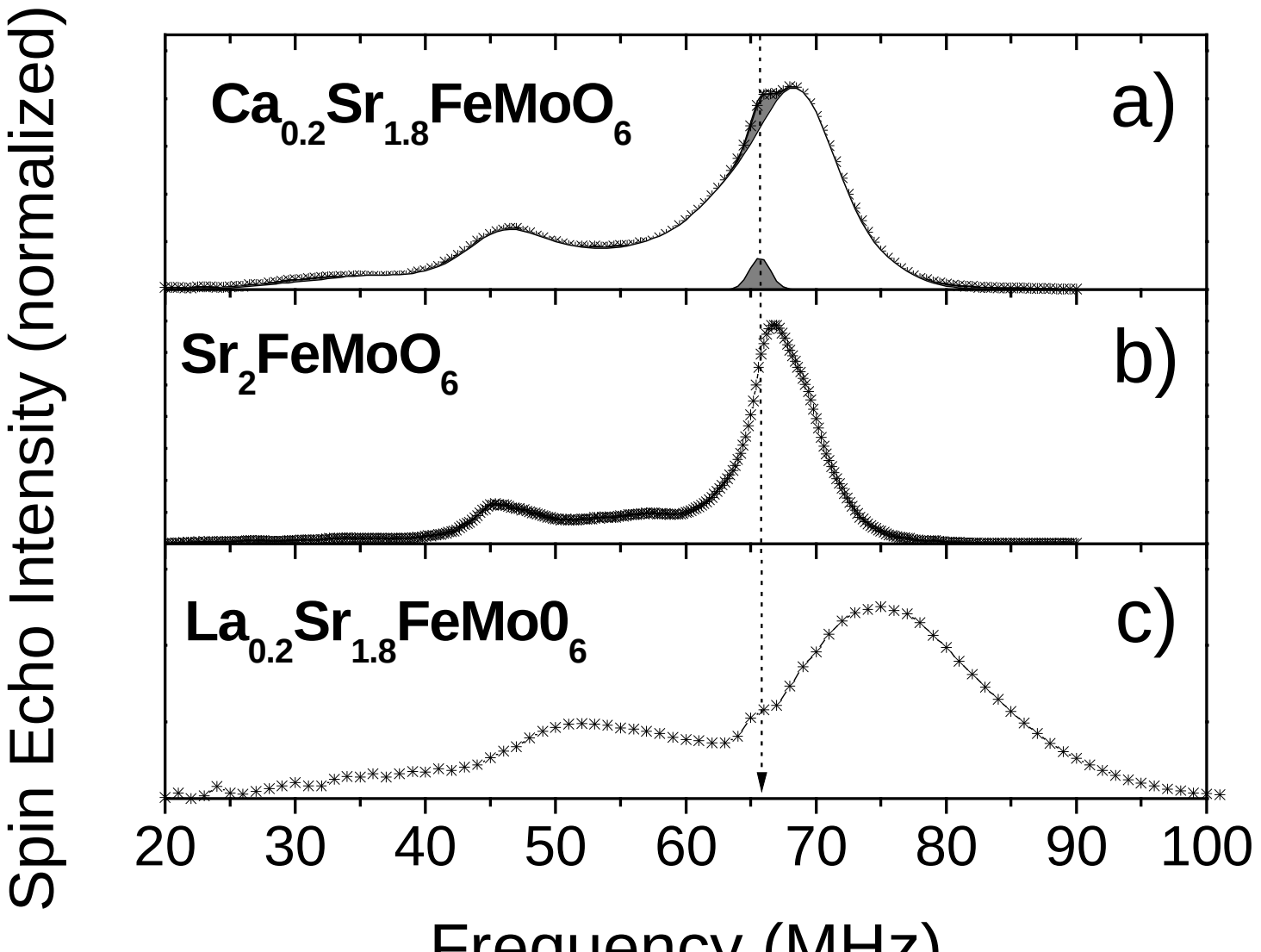

Fig.2, M. Wojcik et al

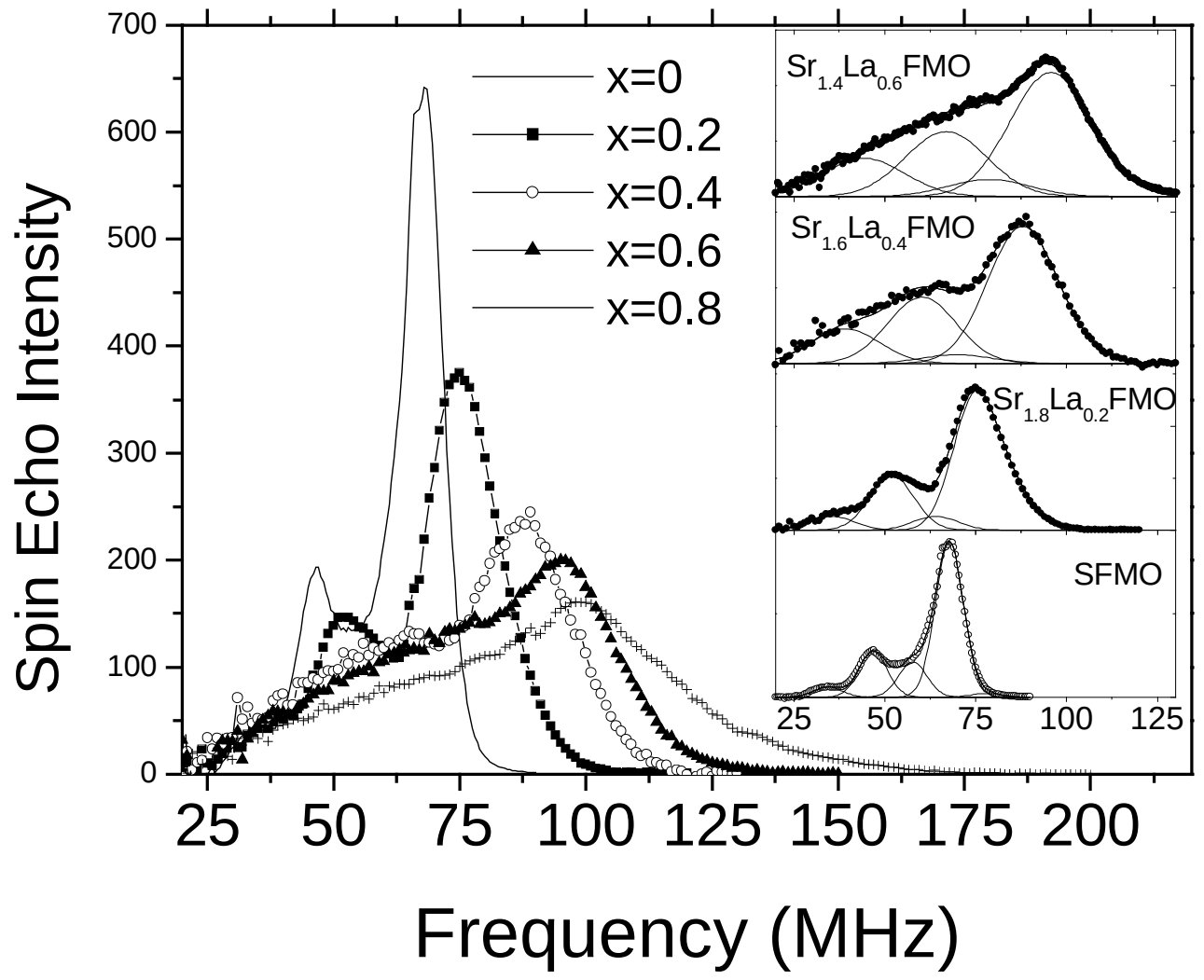

Fig.3, M. Wojcik et al

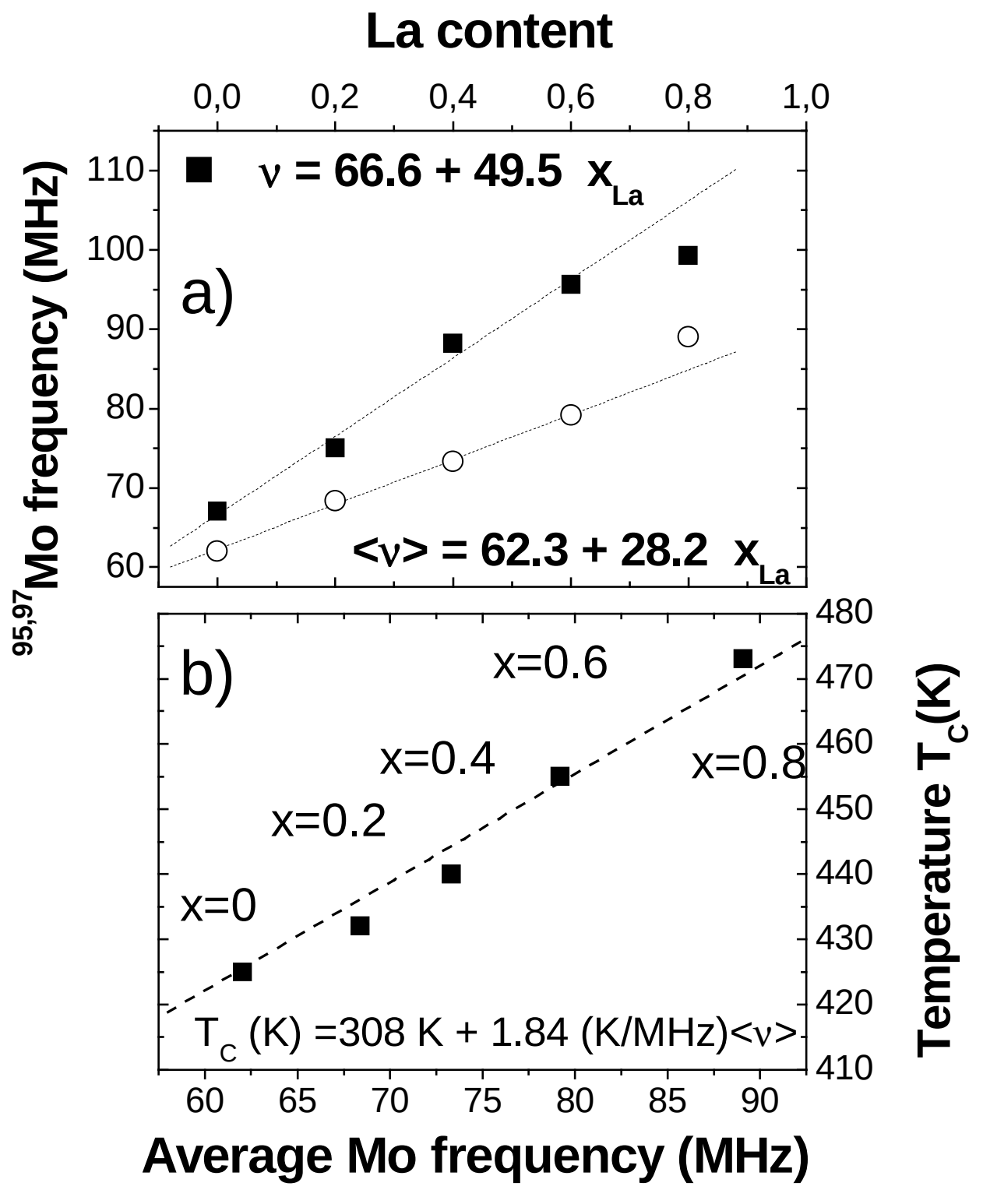

Fig.4, M. Wojcik et al